# Continuous Observation of Interference Fringes from Bose Condensates

J. I. Cirac,[1] C.W. Gardiner,[2] M. Naraschewski,[3,4] and P. Zoller[5]

[1] *Departamento de Fisica Aplicada, Universidad de Castilla-La Mancha, 13071 Ciudad Real, Spain*
[2] *Physics Department, Victoria University, Wellington, New Zealand*
[3] *Max–Planck–Institut für Quantenoptik, Hans–Kopfermann–Straße 1, D-85748 Garching, Germany*
[4] *Sektion Physik, Universität München, Theresienstraße 37, D-80333 München, Germany*
[5] *Institut für Theoretische Physik, Universität Innsbruck, Technikerstraße 25, A-6020 Innsbruck, Austria*

(June 3, 1996)

We use continuous measurement theory to describe the evolution of two Bose condensates in an interference experiment. It is shown how in a single run the system evolves into a state with a fixed relative phase, without violating particle number conservation.

Recent observation of Bose–Einstein Condensation (BEC) [1–3] has initiated theoretical discussions regarding the properties of Bose condensates [4]. Of particular interest have been questions related to the phase of the condensate [5]. The assumption of such a phase as a result of a broken gauge symmetry allows for a natural explanation of many physical phenomena, but also implies that the state of the condensate is a linear superposition of states with different particle number. However, in second quantized formalism of nonrelativistic Quantum Mechanics, as is usually employed to describe BEC, all observables commute with the atomic number operator $\hat{N}$ which thus plays the role of a superselection rule. As a consequence, starting from a state with fixed atomic number or a mixed state, which is diagonal in the atomic number basis, no atomic superpositions and coherences will develop [6].

In order to resolve this seeming contradiction, we use in this paper the language of continuous measurement theory [7] to describe a single realization of an interference experiment between two independent condensates. We will discuss *how the state of the two condensates evolves as atoms are detected*. In particular, from our analysis it follows how a *state of well defined relative phase* builds up dynamically in a single experimental run as a consequence of the von Neuman projection postulate of quantum theory. We emphasize that our description does not contradict atomic number superselection rules, and allows one to understand the coexistence of both particle number conservation and the phase of a condensate in a general situation. The problem addressed in the present paper is related to very recent work by Javanainen and Yoo [8], and Naraschewski *et al.* [9], where it is shown that two independent Bose condensates prepared in Fock states may form a measurable interference pattern. However, here we will show how the interference pattern is formed dynamically, and, what is more important, how the state of the two condensates collapses during the process of detection of the many individual atoms.

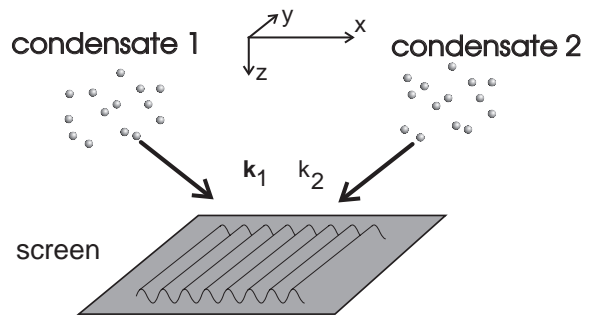

FIG. 1. Schematic setup of interference of two Bose condensates.

We consider the situation depicted in Fig. 1: two statistically *independent* Bose condensates A and B of identical atoms are dropped and detected by atomic counters in the $z = 0$ plane. For simplicity, we assume that all the particles in the first (second) condensate occupy the same mode, which is a plane wave of wavevector $\vec{k}^a$ ($\vec{k}^b$). We also assume that $\vec{k}^a_{y,z} = \vec{k}^b_{y,z}$ and $\vec{k}^a_x = -\vec{k}^b_x$, that is, they have opposite momenta in the $x$ direction [10]. The initial density operator describing the state of the condensates can be written as $\rho_0 = \rho^{(a)} \otimes \rho^{(b)}$. Consistent with the atomic number superselection rule we require $\rho^{(j)}$ to be diagonal in the atomic particle numbers,

$$\rho^{(j)} = \sum_{n=0}^{\infty} p_n^{(j)} |n,j\rangle\langle n,j| \qquad (j=a,b) \tag{1}$$

This includes as a special case Fock states $\rho^{(j)} = |n,j\rangle\langle n,j|$, i.e. states with a fixed atom number (for the existence of interference fringes in single realization of experiments with Fock states see Refs. [8,9]). The derivations which follow are considerably simplified [11] if we assume for the initial states of the condensates an ensemble of phase averaged coherent states with fixed amplitudes $a$ and $b$, respectively,

$$\rho^{(a)} = \int_0^{2\pi} \frac{d\phi_a}{2\pi} |ae^{i\phi_a}\rangle\langle ae^{i\phi_a}|, \tag{2a}$$

$$\rho^{(b)} = \int_0^{2\pi} \frac{d\phi_b}{2\pi} |be^{i\phi_b}\rangle\langle be^{i\phi_b}| \tag{2b}$$

which correspond to Poissonian atomic number distributions $p_n^{(j)} = e^{-j^2} j^{2n}/n!$, $(j = a, b)$.

As the atoms are detected, the number of atoms in each condensate decreases. Due to the fact that the atoms are identical, when one atom is detected one cannot determine whether it belonged to the first or the second condensate, which gives rise to interference. In other words, the number of atoms in each of the condensates is not conserved. However, the total number of atoms must be conserved, and thus the density operator describing the whole system must remain diagonal in the total number. It is therefore convenient to define

$$\hat{R}_{ab}(\phi) = \int_0^{2\pi} \frac{d\phi_a}{2\pi} |ae^{i\phi_a}\rangle\langle ae^{i\phi_a}| \otimes |be^{i(\phi_a-\phi)}\rangle\langle be^{i(\phi_a-\phi)}| \quad (3)$$

as the density operator of a state with fixed relative phase $\phi$. Note that this state is diagonal in the total atom number (consistent with overall number conservation), but is an entangled state with respect to the particles in both condensates [12]. The *initial state* (2) can be then rewritten as

$$\rho_0 = \int_0^{2\pi} \frac{d\phi}{2\pi} \hat{R}_{ab}(\phi) \equiv \int_0^{2\pi} \frac{d\phi}{2\pi} f_{ab}^{(0)}(\phi) \hat{R}_{ab}(\phi) \quad (4)$$

which corresponds to a state with *uniformly distributed random relative phase* $(f_{ab}^{(0)}(\phi) = 1)$.

In order to describe the evolution of the density operator as the atoms are detected at the $z = 0$ plane we use a simple continuous measurement model based on the master equation [13]

$$\dot{\rho} = -i\omega[\hat{a}^\dagger\hat{a} + \hat{b}^\dagger\hat{b}, \rho] + \kappa(2\hat{a}\rho\hat{a}^\dagger - \hat{a}^\dagger\hat{a}\rho - \rho\hat{a}^\dagger\hat{a}) \quad (5)$$
$$+ \kappa(2\hat{b}\rho\hat{b}^\dagger - \hat{b}^\dagger\hat{b}\rho - \rho\hat{b}^\dagger\hat{b}),$$

where $\hat{a}$ and $\hat{b}$ are annihilation operators for particles in the first and second condensate, respectively. The evolution frequency $\omega$ and the loss rate due to detection $\kappa$ are assumed to be equal for the two modes, in accordance with their corresponding momenta $\vec{k}^a$ and $\vec{k}^b$. Note that, according to this master equation, the density operator does not develop atomic superpositions nor coherences. The relationship between this master equation and the physical situation we are considering becomes apparent if we rewrite (5) as

$$\dot{\rho} = -iH_{\text{eff}}\rho + i\rho H_{\text{eff}}^\dagger + \int_0^{2\pi} d\phi \mathcal{J}_\phi \rho, \quad (6)$$

where

$$\mathcal{J}_\phi \rho = 2\kappa(\hat{a} + \hat{b}e^{i\phi})\rho(\hat{a} + \hat{b}e^{i\phi})^\dagger, \quad \phi \in [0, 2\pi), \quad (7a)$$
$$H_{\text{eff}} = (\omega - i\kappa)(\hat{a}^\dagger\hat{a} + \hat{b}^\dagger\hat{b}), \quad (7b)$$

are the (nonhermitian) effective Hamiltonian, and recycling superoperator, respectively. According to continuous measurement theory [7], the interpretation of (6) in the present situations is as follows: a detection of an atom at the position $x_0$ is associated with the action of the jump superoperator as given in (7a),

$$\rho_{t_N+\delta t} \propto \mathcal{J}_\phi \rho_{t_N} \quad (8)$$

with $\phi = [k_x^b - k_x^a]x_0$. That is, an atom is removed from the first *or* the second condensate, and the position of the atom is associated with a relative phase $\phi$ of the plane waves. The time evolution between two subsequent detections is governed by

$$\rho_{t_{N+1}} \propto e^{-iH_{\text{eff}}(t_{N+1}-t_N)} \rho_{t_N} e^{iH_{\text{eff}}^\dagger(t_{N+1}-t_N)}. \quad (9)$$

Let us now follow the time evolution of the atomic density operator in a time interval $[0, t]$ for a sequence of atom detections: we assume that the first atom is detected at the position corresponding to $\phi_1$ and at time $t_1$, the second atom at $\phi_2$, $t_2$ etc. and the $N$-th atom at $\phi_N$, $t_N$. In the present case, starting from the initial state (4) and according to (8) and (9), this sequence of $N$ detections prepares at time $t$ the density operator

$$\rho_t = C \int_0^{2\pi} \frac{d\phi}{2\pi} \prod_{k=1}^{N} |a + be^{i(\phi_k-\phi)}|^2 \hat{R}_{a_tb_t}(\phi) \quad (10)$$
$$\equiv \int_0^{2\pi} \frac{d\phi}{2\pi} f_{ab}^{(N)}(\phi) \hat{R}_{a_tb_t}(\phi)$$

We identify $f_{ab}^{(N)}(\phi) \propto \prod_{k=1}^{N} |a + be^{i(\phi_k-\phi)}|^2$ with the distribution of the relative phase of a state with decaying coherent amplitudes $a_t = ae^{-\kappa t}$, $b_t = be^{-\kappa t}$, and $C$ is a time–independent normalization constant.

The simplicity of Eq. (10) is due to the properties of coherent states: Firstly, they remain coherent but with a decaying amplitude under the time evolution with the effective Hamiltonian $H_{\text{eff}}$, i.e. $e^{-iH_{\text{eff}}t}|a\rangle|b\rangle \propto |a_t\rangle|b_t\rangle$. Secondly, they are eigenstates of the annihilation operator and therefore also eigenstates of the jump superoperators:

$$\mathcal{J}_\psi \hat{R}_{ab}(\phi) = |a + be^{i(\psi-\phi)}|^2 \hat{R}_{ab}(\phi). \quad (11)$$

Note that $f_{ab}^{(N)}(\phi)$ is independent of the jump times $t_k$, which is a consequence of our assumption of equal damping constants, i.e., equal transverse momenta of the two modes.

The probability for observing the next count at position corresponding to the angle $\psi$ is thus given by

$$P(\psi) \propto \int_0^{2\pi} \frac{d\phi}{2\pi} |a + be^{i(\psi-\phi)}|^2 f_{ab}^{(N)}(\phi). \quad (12)$$

Suppose now that that in a particular run an atom is counted at $\phi_{N+1}$. The state immediately after the $N+1$-th detection is

$$\rho_{t+dt} \propto \int_0^{2\pi} \frac{d\phi}{2\pi} |a + be^{i(\phi_{N+1}-\phi)}|^2 f_{ab}^{(N)}(\phi) \hat{R}_{a_tb_t}(\phi), \quad (13)$$

i.e., the relative phase distribution changes according to the map $f_{ab}^{(N+1)}(\phi) \to |a + be^{i(\phi_{N+1}-\phi)}|^2 f_{ab}^{(N)}(\phi)$ (except for a normalization constant). Initially, before the first atom is detected, the position (relative phase) distribution is uniform and $P(\psi) = \text{const}$. The (random) position of the first and subsequent detected atoms gradually determines the relative phase of the condensates, thereby breaking the initial symmetry of the distribution.

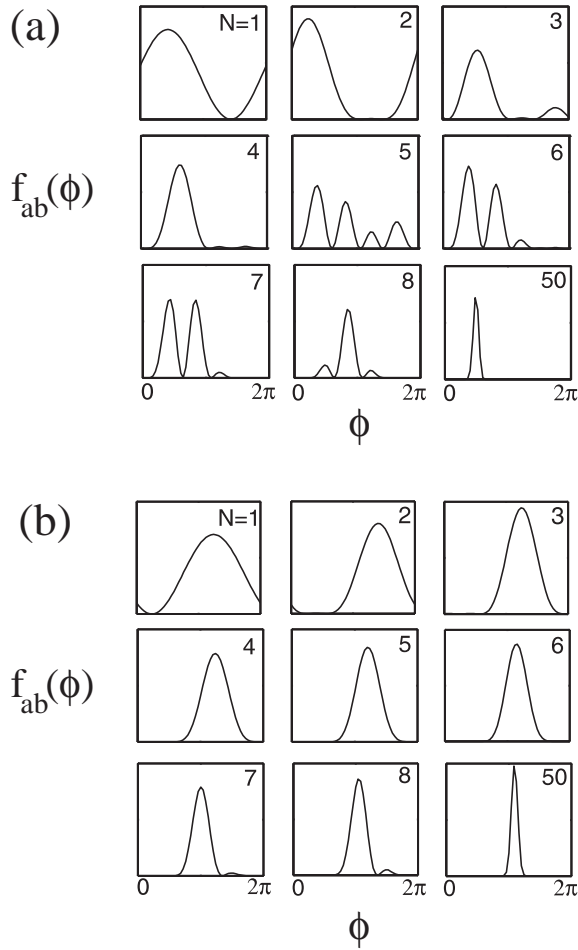

FIG. 2. Evolution of the distribution of the relative phase $f_{ab}^{(N)}(\phi)$ as a function of the number of detected atoms $N = 1, 2, \ldots, 8$ and $N = 50$ for two sample runs. Note, that the scaling of the y-axis is arbitrary.

Figures 2 and 3 give two examples of trajectories which were obtained by simulating a sequence of counts assuming equal initial amplitude $a = b$ (note that in that case the result is independent of the specific value of $a = b$). In Fig. 2(a,b) we have plotted the evolution of the relative phase distribution $f_{ab}^{(N)}(\phi)$ after $N = 1, \ldots, 8$ and after the 50-th detected atom. Typically, the phase distribution is well-peaked after the first few counts. Asymptotically the width of the distribution scales as $\Delta\phi \propto 1/\sqrt{N}$, and $\rho_t$ approaches a state with a well defined relative phase. Of course, the value of this phase varies randomly from run to run. If we average over many realizations, the interference pattern disappears, as expected from master equation (5). Figure 3 shows a simulation of the interference pattern in the $x$-$y$ plane of a single realization of the experiment for three different number of detections $N = 20, 200$ and $1000$ (a,b,c).

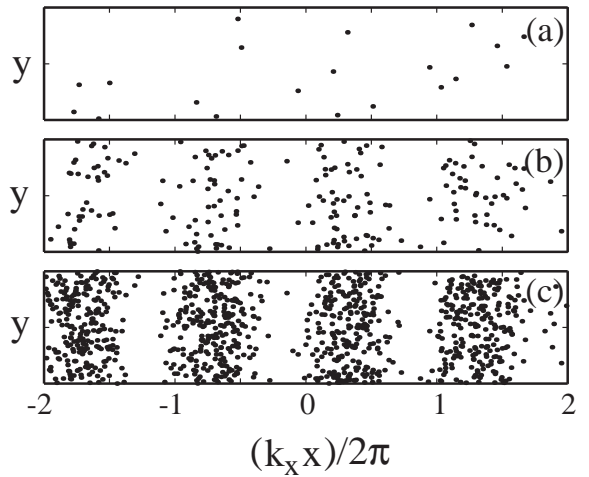

FIG. 3. Buildup of the interference pattern for $N = 20, 200, 1000$ atoms corresponding to (a), (b) and (c), respectively.

Note that the state of relative phase $\hat{R}(\phi)$ is indeed a fixed point of the map defined by the simulation procedure, since it is reproduced by any jump operator $\mathcal{J}_\psi$, $\mathcal{J}_\psi \hat{R}_{ab}(\phi)/\text{Tr}\{\mathcal{J}_\psi \hat{R}_{ab}(\phi)\} = \hat{R}_{ab}(\phi)$, i.e., $\hat{R}_{ab}(\phi)$ does not evolve apart from a decay of its amplitude. The distribution of atoms in the screen corresponding $\hat{R}(\phi)$ is given by

$$P(\psi) \propto |a + be^{i(\psi-\phi)}| = a^2 + b^2 + 2ab\cos(\psi-\phi) \quad (14)$$

and agrees perfectly with the interference pattern that one would expect intuitively from a state of fixed relative phase $\phi$.

An interference experiment of the type described here prepares a state of the form (3) with a fixed relative phase between the two condensates. Note that this state is diagonal in the total number of particles, i.e. does not violate superselection rules, and therefore it gives a consistent description of the experiment. Furthermore, the interference fringes for any possible outcome $(\phi_1, \phi_2, \ldots, \phi_N)$ are completely consistent with the outcome produced if one had initially two coherent states $|a\rangle$ and $|be^{i\phi}\rangle$, with a phase difference $\phi$ coinciding with the asymptotic $(N \to \infty)$ phase difference of $R(\phi)$. This means that for the description of *any* interference experiment with Bose condensates, one can simply assume that the initial state of each of the condensates *is a coherent state with a random phase*. Despite the fact that this state does not comply with the particle superselection rule, the results predicted in any experiment will be the same as those given by a consistent theory such as the one presented here.

The results obtained above can be also explained and generalized using the coherent state representation. In order to do that, we use the following property of Quantum Mechanics: if the state of a system can be written in a diagonal form with respect to a set of states $|\alpha_i\rangle$ (that do not need to be orthogonal), i.e. $\rho = \sum_i P_i |\alpha_i\rangle\langle\alpha_i|$, then a single realization of a given experiment must yield the same result as if the state of the system is one of these states $|\alpha_i\rangle$. Obviously, the probability of obtaining the result corresponding to $|\alpha_i\rangle$ is $P_i$. In our case, we can expand (a large class of) states that are diagonal in the Fock basis representation in terms of coherent states [14]

$$\rho = \int da P(a) \int_0^{2\pi} \frac{d\phi_a}{2\pi} |ae^{i\phi_a}\rangle\langle ae^{i\phi_a}|, \qquad (15)$$

and in particular the states (2). Consequently, a single realization of any experiment must give a result compatible with at least one of the particular coherent states $|ae^{i\phi_a}\rangle$, which leads to the same conclusion as in the preceding paragraph. This interpretation can also be used to show that two condenstates in Fock states may lead to interferences [8,9]. For any physical state of the condensate diagonal in the atomic number, we can write (15) or (1) (both are equivalent [15]). Thus, a single realization of an experiment has to be compatible with both Fock and coherent states. Since for two inital coherent states the probability distribution is of the form (14) (the coherent states are fixed points of any jump operator $\mathcal{J}_\psi$), for two Fock states it must have the same form.

As a final comment, we note that the above discussion implies the existence of a *phase standard* for the relative phase of between Bose condensates. Let us assume that initially we have three condensates A, B and C which have been prepared in the product state $\rho_0 = \rho^{(a)} \otimes \rho^{(b)} \otimes \rho^{(c)}$ corresponding to an ensemble of coherent states $|ae^{i\phi_a}\rangle |be^{i\phi_b}\rangle |ce^{i\phi_c}\rangle$ with uniformly distributed random phases $\phi_a, \phi_b,$ and $\phi_c$. First, observation of intereferences between A and B in a single run will fix the relative phase between A and B, $\alpha = \phi_a - \phi_b$, and thus prepare the state $|ae^{i\phi_a}\rangle |be^{i\phi_a - \alpha}\rangle |ce^{i\phi_c}\rangle$ with $\phi_a, \phi_c$ random. Second, observation of interference between B and C will prepare a state $|ae^{i\phi_a}\rangle |be^{i\phi_a - \alpha}\rangle |ce^{i\phi_a - \beta}\rangle$ with $\phi_a$ random. From this we can now *predict* the phase which will be observed in an interference experiment between A and C, $\gamma = \alpha + \beta$, at least within the uncertainty of our knowledge of the phases $\alpha$ and $\beta$.

In summary, using a description based on continuous measurement theory we have shown how the dynamics in a single run of an interference experiment between two Bose condensates prepares a state with relative fixed phase. Finally, we emphasize that our discussion explicitly assumes that time scale of phase fluctuations due to dynamical properties of the Bose condensate [17] is much slower than the observation times necessary to build up an interference pattern.

After this work was completed we have become aware of two related discussion of the phase of Bose condensate and its measurement by J. Dalibard and K. Mølmer [16].

This work was supported by the Austrian Science Foundation, the Marsden Fund contract GDN-501 under the auspices of the Royal Society of New Zealand, and the Deutsche Forschungsgemeinschaft.

[1] M.H. Anderson *et al.* Science **269**, 189 (1995).
[2] C.C. Bradley, C.A. Sacket, JJ. Tollet and R. Hulet, Phys. Rev. Lett. **75**, 1687 (1995).
[3] K.B. Davis *et al.* Phys. Rev. Lett. **75**, 3969 (1995).
[4] For a review see Bose-Einstein Condensation, edited by A. Griffin, D.W. Snoke and S. Stringari (Cambridge University Press, 1995).
[5] A.J. Leggett in Ref. [4]
[6] An analogous situation exists in laser theory where the number of atomic excitations plus the number of photons is conserved in the rotating wave approximation. Thus for an intially empty cavity and a mixed state of atomic excitations the reduced density operator of the photon field will be diagonal in the photon number basis.
[7] For are review and a complete list of references see P. Zoller and C. W. Gardiner in *Quantum fluctuations*, Les Houches, eds. E. Giacobino *et al.* (Elsevier, NY, in press).
[8] J. Javanainen and S. M. Yoo, Phys. Rev. Lett. **76** , 161 (1996).
[9] M. Naraschewski, J. I. Cirac, H. Wallis, A. Schenzle, and P. Zoller, Phys. Rev. A, (to be published).
[10] For simplicity we ignore the effects of gravity.
[11] This assumption is not essential. Employing a *P*-representation Eq. (2), the following derivations are readily generalized to arbitrary initial states.
[12] This becomes apparent if we rewrite (3) as

$$\hat{R}_{ab}(\phi) = e^{-(a^2+b^2)} \sum_{N=0}^{\infty} \frac{(a^2+b^2)^N}{N!} |\Psi_N\rangle\langle\Psi_N|,$$

where

$$|\Psi_N\rangle = (a^2+b^2)^{-\frac{N}{2}} \sum_{n=0}^{N} \binom{N}{n}^{\frac{1}{2}} a^{(N-n)} b^n e^{-in\phi} |(N-n)\rangle_a |n\rangle_b,$$

is a state with $N$ particles.
[13] M. Naraschewski, diploma thesis, Universität München (1993).
[14] C. W. Gardiner, *Quantum Noise* (Springer-Verlag, Berlin, 1991).
[15] The coherent state is an overcomplete basis, which allows to write a given density operator in a diagoanal form in two different sets, the coherent states and the Fock states.
[16] Talks given by J. Dalibard and K. Möllmer, Conference on *Collective Effects in Ultrcold Atomic Gases*, Les Houches (April 1 - 5, 1996).
[17] M. Lewenstein, private communications.